# 2D nano-granularity of the oxygen chains in the $YBa_2Cu_3O_{6.33}$ superconductor

G. Campi[1], A. Ricci[2], N. Poccia[3] and A. Bianconi[1,3]

[1]*Institute of Crystallography, CNR, via Salaria Km 29.300, Monterotondo Roma, I-00015, Italy*
[2]*Deutsches Elektronen-Synchrotron DESY, Notkestraße 85, D-22607 Hamburg, Germany*
[3]*RICMASS Rome International Center for Materials Science Superstripes, via dei Sabelli 119A, 00185 Roma, Italy*

*Corresponding author*: gaetano.campi@ic.cnr.it

**Abstract** The organization of dopants in high temperature superconductors provides complex topological geometries that controls superconducting properties. This makes the study of dopants spatial distribution of fundamental importance. The mobile oxygen ions, y, in the $CuO_2$ plane of $YBa_2Cu_3O_{6+y}$ (0.33<y<0.67) form ordered chains which greatly affect the transport properties of the material. Here we visualize and characterize the 2D spatial organization of these oxygen chains using scanning micro X-ray diffraction measurements in transmission mode on a thin single crystal slab with y=0.33 ($T_c$=7 K) near the critical doping for the insulator-to-metal transition. We show the typical landscape of percolation made of a granular spatial pattern due the oxygen chains segregating in quasi-one-dimensional needles of Ortho II (O-II) phase embedded in an insulating matrix with low density of disordered oxygen interstitials.

New and complex functional materials are characterized by structural heterogeneity giving rise to textures and granular patterns at different scale lengths. The emergence of multiscale patterns, from the microscale down to the nanoscale, gives new physical-chemical properties in several class of materials belonging to different research field such as chemistry [1, 2], material engineering [3] and biomedicine [4-6], just to mention a few.

High temperature superconductors, constitute an intriguing class of materials where the separation and competition of multiple phases at nanoscopic scale occurs [7-11]. These inhomogeneous phases are due to the ordering of dopants and competition between short range charge, spin and orbital density wave order (CDW, SDW, ODW) inhomogeneity. As a result, new complex geometries emerge, where quantum coherence develops [12-14]. Thus, the study of local features by using high resolution experimental probes assumes a paramount importance for the a deeper understanding of new complex and heterogeneous materials. Local probes such as X ray Absorption experimental methods (XANES and EXAFS) and high resolution X ray and neutron diffraction have been used for investigating structural inhomogeneity in various systems such as diborides [15-17] and cuprates [18-20]. In the last years, the possibility to focus X ray beam onto micrometric and sub-micrometric areas, has allowed to probe the local structure with high resolution in the real space. More specifically, for example, microbeams have been used for mapping weak diffuse scattering in the real space in Scanning micro X Ray Diffraction (SµXRD) measurements. The collected data, treated by spatial statistical tools, have shown the formation of fractal patterns of interstitial oxygens ordered domains in LCO [21], intrinsic phase separation in doped iron-chalcogenides [22, 23] and planar symmetry breaking in Bismuthates [24]. More recently, it's has been demonstrated that the quenched disorder due to dopants is spatially anticorrelated with electronic textures and/or local lattice distortions in Hg1201 [12, 13] and LCO [25]. Thus, the oxygen striped domains are candidate to provide the space where superconductivity arises. This makes the study of quenched disorder and its spatial distribution of fundamental importance.

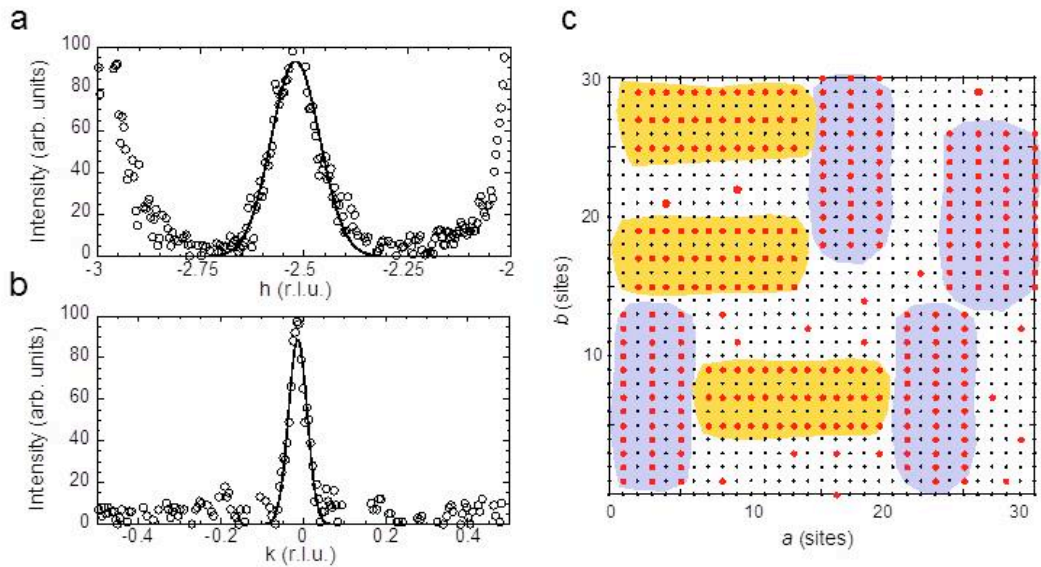

**Figure 1**: (open circles) X-ray diffraction profile of the O-II superlattice with wavevector $q_{O-II}$ = -2.5h, along (a) a* and (b) b* directions. Continuous lines indicate the Gaussian fits. (c) (red full circles) Oxygen defects Oi and (black dots) Cu sites forming the basal $CuO_2$ planes. The Ortho II phase (shadowed areas) made of horizontal and vertical Oi chains are embedded in domains with disordered poor oxygens

$YBa_2Cu_3O_{6+y}$ (YBCO) is one of the most studied high temperature superconductors due to its simple synthesis route and because it was the first superconductor to be discovered with $T_c$ above the liquid nitrogen temperature. Here oxygen dopants get ordered in Cu-O chains that attract electrons from the $CuO_2$ layers. This ordering gives rise to "guest superstructures" detected as satellite peaks in diffraction patterns [26-30]. The spatial distribution of different superstructures have been investigated by us using X ray diffraction in reflection mode, probing the vertical plane ac [27, 30].

In this work, we employed X-ray micro-diffraction in transmission mode on a thin slab of $YB_2C_3O_{6.33}$ with the tetragonal-orthorhombic phase boundary, for visualizing the oxygen chains arrangement on the basal $CuO_2$ planes. At this oxygen concentration superconductivity coexists with the antiferromagnetism, although both are strongly suppressed. The $YBa_2Cu_3O_{6.33}$ single crystals were grown in yttrium-stabilized zirconia crucibles by a flux growth method using chemicals of 99.999% purity for $Y_2O_3$ and CuO, and 99.997% for $BaCO_3$. The impurity level of the crystals has been analyzed by inductively coupled plasma mass spectroscopy. The Zr content of the crystals was found to be less than 10 ppm by weight. The major impurities were Al, Fe, and Zn, the sum of which amounts to less than 0.2% atom per unit cell. The oxygen composition of the crystals was changed by use of gas-volumetric equipment. The technique allows to determine the oxygen composition with an accuracy better than $\Delta x=0.02$. The superconducting critical temperature is found to be $T_c=7$ K.

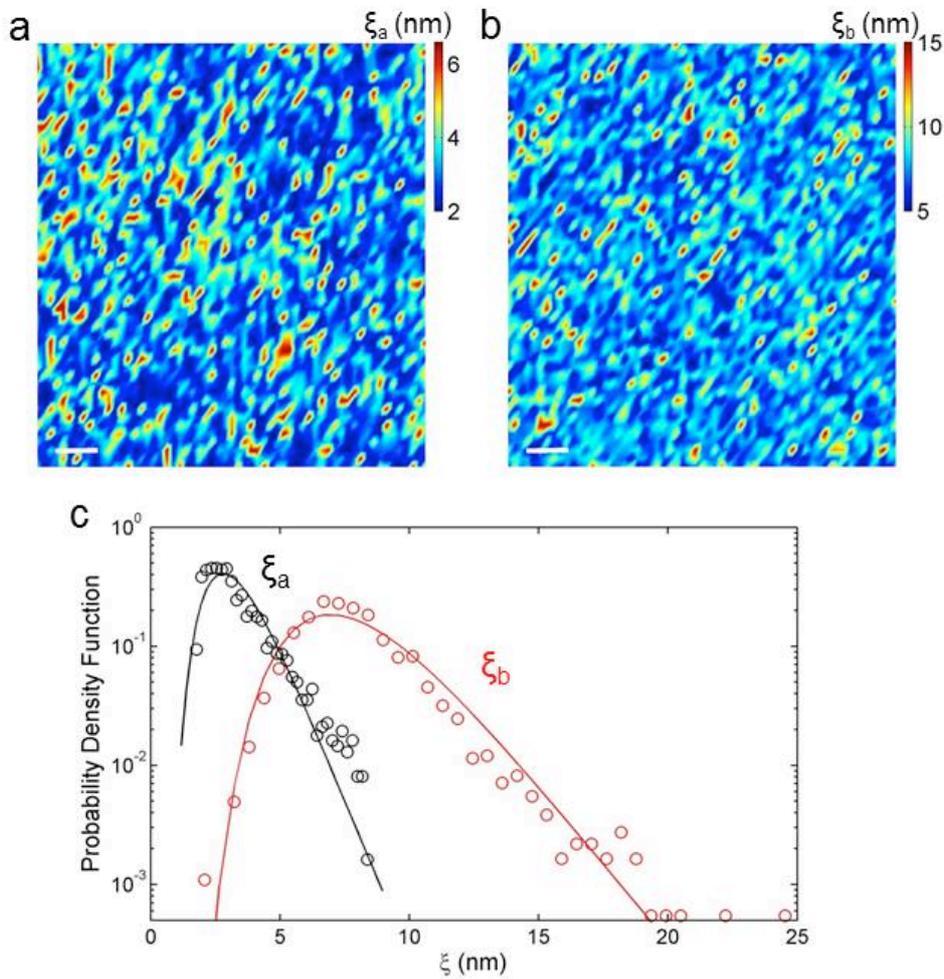

**Figure 2**: Color map of OII correlation length along the (a) a and (b) b directions: the O-II superstructure in plane size domains are plotted as a function of the illuminated spot position XY in the sample surface, where X and Y directions in the image correspond to the a and b crystallographic axis of the sample. The white bar indicates 25 μm length scale on the sample surface. (c) Semi-log plot of spatial distribution of domain size along a-axis and b-axis of OII superlattices. We have used a lognormal function (continuous lines) to model the Probability Density Functions.

Standard X ray diffraction measurements were performed at XRD beamline at ELETTRA, in Trieste, Italy [31] using a photon energy of 20 KeV and a Mar-Research 165 mm Charge Coupled Device (CCD) camera. The unit cell of $YBa_2Cu_3O_{6.33}$ single crystal has a=3.851(4) Å, c=11.78(5) Å, in the I4mm space group. We found superlattice reflections associated to the ortho-II phase, located at positions $q_{O-II}$=(*h*±1/2, *k*±1/2, *l*) with *h*, *k*, and *l* as integers. In **Figure 1a** and **Figure 1b** we show the diffraction profiles along a* and b* direction, respectively, of the (-2.5,0,0) streak satellite reflection, measured at XRD beamline at ELETTRA, fitted by Gaussian line profiles. A pictorial view of the phase separation due to the arrangement of oxygen ions in linear chains with periodicity of 2 unit cells along a and b directions, is shown in **Figure 1c**. The Oi rich domains (with Oi chain fragments) form nanoscale O-II phase coexisting with domains with poor and disordered oxygen.

The spatial distribution of these Oi rich domains has been studied by Scanning micro X ray Diffraction (SµXRD) measurements in transmission mode performed at the ID13 beamline of the European Synchrotron Radiation Facility (ESRF), Grenoble, France. We used a monochromatic X-ray beam with an energy of 14 KeV ($\Delta E/E = 10^{-4}$) focused by Kirkpatrick Baez (KB) mirrors to a 1 µm spot size on the sample. A 16 bit two-dimensional Fast Readout Low Noise charged coupled device (FReLoN CCD) detector with 2048x2048 pixels of 51x51 µm$^2$ was used, binned to 512x512 pixels. Diffraction images were obtained after correcting the 2D images for dark noise, flat field, spatial distortion. This satellite reflections was then measured at each point of the sample reached by the x-y translator with sub-micron resolution in order to visualize the point to point spatial variation of both intensity and size of the "Ortho-II" domains. The correlation lengths along the *a* and *b* directions, $\xi_a$ and $\xi_b$, of the Ortho-II domains, have been derived from the measured FWHM via standard methods of diffraction.

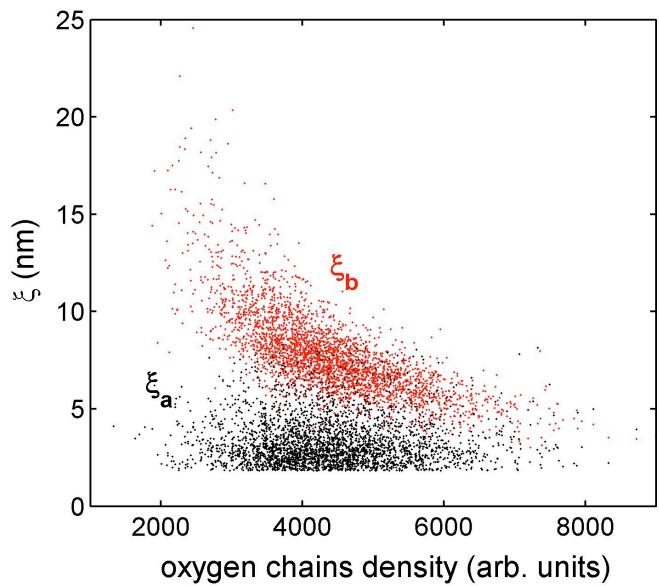

**Figure 3**: Scatter plot of correlation lengths $\xi_a$ and $\xi_b$ as a function of streak intensity giving the density of Oi chains.

The spatial distributions of $\xi_a$ and $\xi_b$ are shown by the maps in Figure 2a and Figure 2b, respectively. We can clearly observe the inhomogeneous character of the size of the ordered domains. In order to characterize this spatial texture, we calculated the Probability Density Function of both $\xi_a$ and $\xi_b$ (see Figure 2c). As previously determined by statistical analysis of domain sizes, some deviations from normal behavior are observed in the right tails of distribution [27-30]. These deviations can be quantified by the distribution skewness, *sk*, giving $sk_a$=1.35 and $sk_b$=28.5 for the Ortho-II domain size along the *a* and *b* directions, respectively. Indeed, the probability density function of $\xi$, shown in Figure 2c appears clearly with a fatter tail in the (long) *b* direction. The correlation lengths range from 2 to 8 nm in the (short) *a* direction and from 4 to 30 nm in the (long) *b* direction. It is well known how deviations from a normal distribution in the right tail of a distribution are related with a more complex behavior in several systems and processes [32]. In this case, the fatter tails suggest a complex morphology for the granular ab surface due to the presence of *few* large domains coexisting with *many* small domains of Oi chains.

This is confirmed by investigating the interplay between the density and size of rich Oi chain domains by the scatter plot of the domain size along *a* and *b* as a function of $\mathbf{q_{O-II}}$ superstructures intensity (**Figure 3**). We note that the size of ordered domains along *b* is anti-correlated with the domains population; in other words, we have few large domains and many small domains along *b*. On the other side, we get a quit random relationship between the domains population and domain size along a.

In summary, we have studied the planar granularity due to the oxygen order in $YBa_2Cu_3O_{6.33}$. We measured the local variations of oxygen ordered domains by analyzing the satellite reflections associated to the Orto II needle-like domains made of oxygen chains running along the a and b directions. Our observations clearly show that chains ordering in YBCO produces a phase separation between 2D ordered and disordered domains with poor oxygen content (y<0.33) with the typical complex topology of a superstripes landscape [33]. The ordered O-II domains form quasi one-dimensional needles with a distribution of size ranging between 2 and 8 nanometers in the transversal direction and between 3 and 30 nm along the needle direction. Finally we would remark that our XRD diffraction experiment show that $YBa_2Cu_3O_{6.33}$ is made of $YBa_2Cu_3O_6$ with a percolating concentration of 66% of $YBa_2Cu_3O_{6.5}$ needles. Therefore it is possible that the spatial topology for the emergence of high temperature superconductivity is given by the interface space between the one dimensional O-II needles with can give a hyperbolic space [12,13].